# The Different Shapes of the LIS Energy Spectra of Cosmic Ray He and C Nuclei Below ~1 GeV/nuc and The Cosmic Ray He/C Nuclei Ratio vs. Energy - V1 Measurements and LBM Propagation Predictions

W.R. Webber

New Mexico State University, Astronomy Department, Las Cruces, NM  88003, USA

**ABSTRACT**

This paper examines the cosmic ray He and C nuclei spectra below ~1 GeV/nuc, as well as the very rapid increase in the He/C ratio below ~100 MeV/nuc, measured by Voyager 1 beyond the heliopause. Using a simple Leaky Box Model (LBM) for galactic propagation we have not been able to simultaneously reproduce the individual He and C nuclei spectra and the large increase in He/C ratio that is observed at low energies. However, using a truncated LBM with different truncation parameters for each nucleus that are related to their rate of energy loss by ionization which is ~$Z^2/A$, these different features can be matched. This suggests that we are observing the effects of the source distribution of cosmic rays in the galaxy on the low energy spectra of cosmic ray nuclei and that there may be a paucity of nearby sources. In this propagation model we start very specific source spectra for He and C which are ~$dj/dP = P^{-2.24}$, the same for each nucleus and also for all rigidities. These source spectra become spectra with spectral indices ~-2.69 at high rigidities for both charges as a result of a rigidity dependence of the diffusion coefficient governing the propagation which is taken to be ~$P^{-0.45}$. This exponent is determined directly from the B/C ratio measured by AMS-2. These propagated $P^{-2.69}$ spectra, when extended to high energies, predict He and C intensities and a He/C ratio that are within $\pm$3-5% of the intensities and ratio recently measured by AMS-2 in the energy range from 10 to 1000 GeV/nuc.

**Introduction**

Detailed spectra are now available for He and heavier nuclei from the Voyager 1 spacecraft at low and intermediate energies below ~1 GeV/nuc (Cummings, et al., 2016; Webber, et al., 2017). The differential intensities of those spectra for various nuclei appear to reach their maximum at ~30 MeV/nuc for He nuclei and at somewhat higher energies for the heavier nuclei such as C and O, up to Fe. The fact that these spectra reach a peak is a product of the acceleration and propagation processes (e.g., Cummings, et al., 2017) and also possibly the nearby source distribution of these galactic cosmic rays. It has only been possible to study these low energy spectra since V1 left the standard heliosphere in August, 2012, and the solar modulation effects on these spectra dropped to ~zero. In a sense a new area of study of galactic cosmic rays was born at that time.

Webber and Higbie, 2015, have noted apparent differences in the low energy He and C spectra measured by Voyager. Webber, et al., 2017, have also recognized that the intensities of He and heavier nuclei at ~100 MeV/nuc measured at V1, when compared with intensities of these nuclei measured at 1.0 GeV/nuc, varied in a systematic Z dependent way. These Z dependent differences could be explained if one assumed, to start with, that the source spectra of the nuclei could be described as:

$$dj/dP = \text{const} \times P^{-2.24}$$

for each nucleus, where P is rigidity. The spectral index is the same for both nuclei and its value of -2.24 is independent of rigidity. Given that these are the source spectra, the Z dependent differences that are actually observed for the measured spectra are mostly determined by the features of the propagation. In this case the path length is the key parameter and it is related to the index of the diffusion coefficient dependence on rigidity which is typically taken to be between 0.33-0.50 above 1 GV. This dependence determines the amount of matter traversed and the amount of energy loss by ionization of the particles which depends on the $Z^2/A$ of the nuclei.

Measurements of the B/C ratio by AMS-2 up to several hundred GeV/nuc (Aguilar, et al., 2016) suggest the rigidity dependence of the diffusion coefficient is ~$P^{-0.45}$ above ~1 GV (Webber and Villa, 2016, 2017) which when added to the source index -2.24, gives a propagated spectral index ~-2.69 at high rigidities which is, indeed, almost exactly the new average high rigidity spectral index ~2.70 measured by AMS-2 for both He and C nuclei (Aguilar, et al., 2015, 2016).

In a Leaky Box Model (LBM) the mean path length, $\lambda(P) \sim \rho v T$, where $\rho$ is the density of the IS medium, v is the particle velocity and T is the particle diffusion lifetime, $T = \ell^2/D(P)$, $\ell =$ the scale size of the trapping region and $D(P)$ is the diffusion coefficient in $cm^2 \cdot sec$.

For energies below ~100 MeV/nuc the situation becomes more complicated. Webber, 2016, has described the observed Z dependence of the intensity maximum of the spectra of He, C and heavier nuclei in terms of the distribution of nearby cosmic ray sources in a truncated LBM. The ionization energy loss of these different species, which is $\sim Z^2/A$, is indeed still a dominant effect at energies below ~100 MeV and is creating a Z dependent depletion of the heavier species. However the differences in spectra measured at Voyager appear to be even more severe than those produced only by ionization loss effects in a LBM, hence the need to modify the actual path length distribution, which in a normal LBM calculation is an exponential at each rigidity (see e.g., Slickheiser and Lerche, 1985).

In this paper we will present calculations of the He and C spectra at low energies including estimates of the values of the peak energies of the differential spectra using a modified exponential PLD and compare these peak energies with those measured for He and C on Voyager.

**The Data**

In Figure 1 we show the data for He and C nuclei from Cummings, et al., 2016, at energies below ~100 MeV/nuc and Webber, et al., 2017, at energies ~1 GeV/nuc on the same graph, with the C nuclei intensities multiplied by a factor of 35.0. It is clear from the graph that there is a strong Z dependent effect on these spectra at energies below ~100 MeV/nuc. This Z dependence is still evident between 100 and 1000 GeV/nuc.

This spectral difference is shown even more clearly in the energy dependence of the He/C ratio shown in Figure 2. The He/C ratio that is measured at higher energies by AMS-2 is also shown in Figure 2. This ratio has a value of about 32.5 at 3 GeV/nuc and this ratio slowly decreases in a methodical way to ~29 at 10 GeV/nuc and then to about 27.5 at 100 GeV/nuc remaining at this value, $\pm 1$, out to ~1000 GeV/nuc. The precision of these values is $\pm$ a few percent.

If the same rigidity dependence of the diffusion coefficient that is determined above a few GeV/nuc (e.g., $P^{0.45}$) extends down to a $P_0$=1.0 GV (~125 MeV/nuc) then the calculated He/C dependence does indeed match the newest Voyager observations at ~100 MeV/nuc and above and also the AMS-2 observations. However, the observed He/C ratio increases much more rapidly at lower energies than the simple LBM model calculations made using the above parameters, which include the full effects of ionization E loss but have a standard exponential PLD for the path length at lower rigidities.

## **Matching The Data at Energies Below ~100 MeV/nuc**

The problem here is to simultaneously match the Voyager data on He and C nuclei spectra individually and also the He/C ratio at low energies. If one assumes a simple LBM calculation in which the path length $\lambda = 20.6 \, \beta \, P^{-0.45}$, and that the rigidity dependence of 0.45 determined at higher rigidities extends down to ~0.562 GV with a break in the rigidity dependence at $P_0$=0.562 GV or below, (a reasonable assumption supported by the work of Ptuskin, et al., 2006), then the choice of $P_0$=0.562 GV does now give a match to the Voyager measured individual He and C spectra down to energies ~20-30 MeV/nuc. It also matches the He/C ratio data above ~100 MeV/nuc but it still does not match at lower energies (see Figure 2). For the choice of a lower value of $P_0$=0.562 GV the calculation still does not fit the measured rapidly increasing the He/C ratio that occurs below 20-30 MeV/nuc. Also the calculated He and C intensities at lower energies now individually become larger than those measured at Voyager.

At this point one considers a modification to the standard LBM known as a truncation of short path lengths introduced many years ago (Garcia-Munoz, et al., 1987, see also Webber, 2016). In this picture the path length P(x) may be described as the sum of two exponential PLD, $\lambda_1$ and $\lambda_2$. The resultant PLD is

$$P(x) = [e^{-x/\lambda_1} + e^{-x/\lambda_2}]$$

where $\lambda_2 \ll \lambda_1$. In a NLBM such as suggested by Cowsik and Wilson, 1975, where some of the material is transversed near the source in the case where $\lambda_2$ is small and may represent the material around the source, $\lambda_2$ may be the same for all charges as it was in the original Cowsik and Wilson model.

If we make $\lambda_2 = 0.04$, then the He spectrum itself is now well fit, but the predicted C spectrum is still above the data and, although the calculated He/C ratio is now somewhat larger below 100 MeV/nuc, it is still below the Voyager observations. However, if we choose $\lambda_2 = 0.04$ for He and 0.12 for C we now obtain good fits (for $P_0=0.562$) to both the individual He and C spectra and also the He/C ratio shown in Figures 1 and 2. The use of different values of $\lambda_2$ for different Z nuclei would be compatible with the idea that the different mean path lengths for different Z in the truncation may be due to ionization energy loss which is $\sim Z^2/A$. For He this quantity equal 1.0 but for C it is 3.0 which is just the ratio of the rates of ionization energy loss, $Z^2/A$, of the two species. A distribution of cosmic ray sources in the galaxy in which there is paucity of nearby sources could produce a similar effect (see e.g., Lezniak and Webber, 1977).

Figure 3 illustrates the differences in the mean path length for smaller path lengths for a pure exponential =8 g/cm$^2$ and for values of $\lambda_2=0.04$ and 0.12.

**The Peaks in the Differential Energy Spectra for He, C and Heavier Nuclei**

The truncated LBM for cosmic ray propagation as described above makes specific predictions for the energy at which the differential spectra of various cosmic ray nuclei reach their peak intensity. In Table 1 we show this calculated energy/nuc, also including Hydrogen, Helium and heavier nuclei such as C, O, Mg, Si and Fe, measured by Voyager. The energy of the maximum intensity increases from ~32 to 96 MeV/nuc for H, He and Fe respectively, for a truncated LBM. We also show calculations for a simple LBM with $P_0=1.0$ GV and also for 0.562 GV, each of which could fit the He or C intensities but not the He/C ratio. In this model the energy of the maximum energy is significantly lower and increases from ~20 to 45 MeV/nuc mainly because of the $Z^2/A$ dependence of ionization energy loss for H, He and Fe respectively in the simple LBM.

In the right hand column of Table I is the estimated energy of the maximum differential intensity from the published measured intensities for the various nuclei in Cummings, et al., 2016. These measured values range from 20-24 MeV/nuc for H and He up to ~60-70 MeV/nuc for C and O nuclei and up to ~100 MeV/nuc for Fe nuclei. These values agree with the truncated LBM predictions for C through Fe nuclei, but are more like the simple LBM calculation for H and He nuclei spectra. We note here that new, soon to be published, data on stopping H and He nuclei in the energy range ~20-72 MeV/nuc may indicate a higher peak energy ~30 MeV/nuc for H and He nuclei.

## **The Predicted He and C Intensities and the He/C Ratio at High Energies and the AMS-2 Measurements**

Voyager 1 measurements of the He and C nuclei intensities and the intensity ratio, He/C, have been made from a few MeV/nuc up to ~1.5 GeV/nuc (Cummings, et al, 2016; Webber, et al., 2017). These measurements have led to the determination of source rigidity spectra ~$P^{-2.24}$ and an interstellar path length $\lambda=20.6\ \beta\ P^{-0.45}$ appropriate to rigidities >0.562 GV as noted above. When extended to higher rigidities these parameters lead to high rigidity spectra of both He and C that become proportional to $P^{-2.69}$.

In the truncated LBM used to make the lower energy calculations, we now predict the intensities of He and C nuclei and the He/C ratio at energies/nuc up to 1 TeV/nuc. The source ratio $^4$He/$^{12}$C for these calculation is taken to be =23.7. These predictions are compared with the very precise measurements of He and C nuclei intensities by the AMS-2 experiment (Aguilar, et al., 2015, 2016) in Table 2. This comparison is made at 5 energies from 10-1000 GeV/nuc where any effects due to solar modulation are small. It is seen that these AMS-2 intensities and He/C ratios are consistent within $\pm$3% with the predictions of this LBM propagation picture at an energy of >3 GeV/nuc. This remarkable agreement continues up to 1000 GeV/nuc where the disagreement is never more than $\pm$5% from the mean at that energy. Note that a difference of $\pm$0.02 in the spectral index of -2.69 that fits to the AMS spectra from ~8 GV to 2000 GV would amount to $\pm$10% difference in intensities over the range of energies used. This difference is outside of the variation of the predicted to measured intensities.

Table III shows the predictions of the truncated LBM for the sums of He and C nuclei from ~7.5 MeV/nuc to 1 TeV/nuc. Note the significant contribution of $^3$He to the total He component, reaching a fraction ~20% at ~1 GeV/nuc and just above. Also note the effect of truncation on the He/C ratio as shown by the shaded regions in Figure 2 at both low and high energies. At 1 TeV/nuc this predicted ratio is 29.2 in agreement with AMS-2 measurements. For a simple LBM with a source ratio =23.7 this calculated ratio would be only 24.3 at 1 TeV/nuc.

At higher energies the path length used in this paper decreases from ~7.0 g/cm$^2$ at 3 GeV/nuc to 4.2 g/cm$^2$ at 10 GeV/nuc, to 1.5 g/cm$^2$ at 100 GeV/nuc and finally to ~0.53 g/cm$^2$ at 1000 GeV/nuc, for a total decrease factor of 13.0 between 3 and 1000 GeV/nuc. These values for the path length are based on studies using the B/C ratio at corresponding energies measured by AMS-2 (Webber and Villa, 2016, 2017). This variation in path length of a factor ~13.0 $\pm$1.0 between ~8 and 2000 GV rigidity leads to an average diffusion coefficient in the propagation program which is ~P$^{0.45\pm0.05}$.

**<u>Summary and Conclusions</u>**

In this paper we have examined the He and C nuclei spectra of galactic cosmic rays below ~1.5 GeV/nuc as well as the very rapid increase in the He/C ratio below ~100 MeV/nuc measured by V1 beyond the heliopause in interstellar space. We also compare these measurements with those from AMS-2 from ~3 GeV/nuc to 1 TeV and with calculations using a truncated LBM for the propagation of cosmic rays in the galaxy.

First we note that, at low energies, it has not been possible to fit both the He and C spectra and the He/C ratio simultaneously using identical source spectra for He and C that are ~P$^{-2.24}$ independent of rigidity (no Break) and a simple LBM for the propagation in which the mean path length is an exponential for all path lengths (e.g., Schlickeiser and Lerche, 1985). The main difficulty is fitting the rapidly increasing He/C ratio below ~100 MeV/nuc.

The inconsistency noted above can be overcome by considering a truncated LBM in which different truncation parameters are used for He and C. If the truncation parameters are chosen to be 0.04 for He and 0.12 for C then both the He and C spectra and the He/C ratio below 100 MeV/nuc can be simultaneously fit. The ratio of 3.0 for the values of the truncation parameter is equal to the ratio of the rate of ionization energy loss between He and C which is ~Z$^2$/A. This suggests that the rapidly increasing He/C ratio that is observed could be a result of a deficiency of low energy particles and that this deficiency is dependent on the range of these particles. This could be due to a distribution of the sources of these particles that is deficient in nearby sources.

The model we use, which has identical source spectra $\sim P^{-2.24}$ but a truncated PLD, leads to specific predictions of the energy at which the maximum in the differential energy spectrum occurs. The maximum energy is found to increase with increasing Z from an energy ~30 MeV/nuc for He, to ~65 MeV/nuc for C and to ~100 MeV/nuc for Fe in rough agreement with the still the quite uncertain values for these quantities obtained to date at V1.

The model we use, again with identical source spectra $\sim P^{-2.24}$ and with the propagation dependent on a path length $\lambda = 20.6\ \beta\ P^{-0.45}$ above $P_0 = 0.562$, predicts nearly identical high rigidity spectra $\sim P^{-2.69}$ for both He and C nuclei. The predicted intensities from this truncated LBM match those for He and C and also the He/C ratio measured by AMS-2 from 3-1000 GeV/nuc, where solar modulation effects are negligible, to within $\pm 3\%$ at 10 GeV/nuc and to within $\pm 5\%$ at 1000 GeV/nuc. The rigidity dependence of $\lambda \sim P^{-0.45}$ used above is derived from a fit to the AMS-2 B/C ratio measured between 3 to 1000 GeV/nuc (Webber and Villa, 2016, 2017) and is accurate to within $\pm 0.05$ in the exponent.

The predictions of the GALPROP-DR model (see Cummings, et al., 2016), also shown in Figures 1 and 2, are significantly different from the observations of the He/C ratio at both low and high energies. At energies less than ~100 MeV/nuc the Voyager measured He/C ratio is under-predicted by ~10-40% by GALPROP, and at higher energies above ~10 GeV/nuc the GALPROP prediction of this ratio diverges from the AMS-2 measurements by an energy dependent factor which would be equivalent to the rigidity dependence of the intensity of one component (He) being different than that of the other (C) by 0.16 in the exponent.

**Acknowledgements:** The authors are grateful to the Voyager team that designed and built the CRS experiment with the hope that one day it would measure the galactic spectra of nuclei and electrons. This includes the present team with Ed Stone as PI, Alan Cummings, Nand Lal and Bryant Heikkila, and to others who are no longer members of the team, F.B. McDonald and R.E. Vogt. Their prescience will not be forgotten. This work has been supported throughout the more than 40 years since the launch of the Voyagers by the JPL.

| TABLE I Comparison of the Calculated and Measured Differential Intensity Maximum of the Spectra of Various Nuclei | | | |
|---|---|---|---|
| Charge | Truncated LBM Intensity MAX (MeV/nuc) | Simple LBM Intensity MAX (MeV/nuc) | Measured by Voyager Intensity MAX (MeV/nuc) |
| H | 33 | 22 | 20-30 |
| He | 32 | 19 | 24-36 |
| C | 64 | 32 | 75-80 |
| O | 68 | 36 | 80-90 |
| Mg | 88 | 38 | 95-105 |
| Si | 88 | 38 | 85-100 |
| Fe | 95 | 45 | 100-125 |

| Energy GeV/nuc | Calculated/Measured by AMS-2 |||
|---|---|---|---|
| | He[+] Calc/Meas | C[+] Calc/Meas | He/C* Ratio Calc/Meas |
| 10 | 12.6/12.6 | 0.418/0.431 | 30.1/29.2 |
| 31.6 | 0.690/0.669 | 2.39/2.32 E-02 | 28.8/28.8 |
| 100 | 3.31/3.11 E-02 | 1.21/1.50 E-03 | 27.3/28.2 |
| 316 | 1.51/1.39 E-03 | 5.39/4.97 E-05 | 28.3/28.0 |
| 1000 | 6.71/6.57 E-05 | 2.27/2.26 E-06 | 29.5/29.1 |

**TABLE II**
**Comparison of the Truncated LBM Calculations And the Measurements of AMS-2**

[+] Intensities are in P/m$^2$·sr·s·MeV/nuc
*He/C Source Ratio =23.7 for calculations

| TABLE III Intensities of He and C Nuclei in Particles/m$^2\cdot$ sr$\cdot$ s$\cdot$ MeV/nuc | | | | |
|---|---|---|---|---|
| Energy MeV/nuc | He3 | He4 | Sum He | Sum C |
| 7.5 | 5.08E-02 | 1.91E+00 | 1.96E+00 | 1.69E-02 |
| 10 | 6.45E-02 | 2.14E+00 | 2.20E+00 | 2.08E-02 |
| 13 | 8.08E-02 | 2.32E+00 | 2.40E+00 | 2.47E-02 |
| 18 | 9.99E-02 | 2.46E+00 | 2.56E+00 | 2.88E-02 |
| 24 | 1.20E-01 | 2.54E+00 | 2.66E+00 | 3.29E-02 |
| 32 | 1.40E-01 | 2.56E+00 | 2.70E+00 | 3.65E-02 |
| 42 | 1.58E-01 | 2.50E+00 | 2.66E+00 | 3.93E-02 |
| 56 | 1.69E-01 | 2.36E+00 | 2.52E+00 | 4.04E-02 |
| 75 | 1.75E-01 | 2.15E+00 | 2.32E+00 | 4.01E-02 |
| 100 | 1.71E-01 | 1.89E+00 | 2.06E+00 | 3.84E-02 |
| 133 | 1.56E-01 | 1.60E+00 | 1.75E+00 | 3.49E-02 |
| 178 | 1.35E-01 | 1.29E+00 | 1.43E+00 | 3.02E-02 |
| 237 | 1.43E-01 | 9.95E-01 | 1.14E+00 | 2.49E-02 |
| 316 | 1.16E-01 | 7.30E-01 | 8.47E-01 | 1.94E-02 |
| 422 | 8.74E-02 | 5.09E-01 | 5.96E-01 | 1.43E-02 |
| 562 | 6.14E-02 | 3.36E-01 | 3.97E-01 | 1.01E-02 |
| 750 | 4.04E-02 | 2.12E-01 | 2.52E-01 | 6.78E-03 |
| 1000 | 2.52E-02 | 1.30E-01 | 1.55E-01 | 4.40E-03 |
| 1334 | 1.53E-02 | 7.85E-02 | 9.38E-02 | 2.73E-03 |
| 1779 | 8.91E-03 | 4.62E-02 | 5.51E-02 | 1.64E-03 |
| 2372 | 4.99E-03 | 2.64E-02 | 3.14E-02 | 9.48E-04 |
| 3163 | 2.69E-03 | 1.47E-02 | 1.74E-02 | 5.35E-04 |
| 4218 | 1.40E-03 | 7.96E-03 | 9.36E-03 | 2.93E-04 |
| 5624 | 7.03E-04 | 4.20E-03 | 4.91E-03 | 1.56E-04 |
| 7500 | 3.42E-04 | 2.17E-03 | 2.52E-03 | 8.12E-05 |
| 10001 | 1.62E-04 | 1.10E-03 | 1.26E-03 | 4.14E-05 |
| 13337 | 7.45E-05 | 5.48E-04 | 6.23E-04 | 2.07E-05 |
| 17785 | 3.36E-05 | 2.69E-04 | 3.03E-04 | 1.02E-05 |
| 23717 | 1.48E-05 | 1.30E-04 | 1.45E-04 | 4.96E-06 |
| 31627 | 6.45E-06 | 6.26E-05 | 6.90E-05 | 2.39E-06 |
| 42176 | 2.77E-06 | 2.98E-05 | 3.26E-05 | 1.14E-06 |
| 56242 | 1.18E-06 | 1.41E-05 | 1.53E-05 | 5.38E-07 |
| 75000 | 4.98E-07 | 6.63E-06 | 7.12E-06 | 2.53E-07 |
| 100014 | 2.09E-07 | 3.11E-06 | 3.31E-06 | 1.18E-07 |
| 133371 | 8.71E-08 | 1.45E-06 | 1.54E-06 | 5.49E-08 |
| 177853 | 3.62E-08 | 6.74E-07 | 7.11E-07 | 2.54E-08 |
| 237171 | 1.50E-08 | 3.13E-07 | 3.28E-07 | 1.17E-08 |
| 316273 | 6.17E-09 | 1.45E-07 | 1.51E-07 | 5.37E-09 |
| 421757 | 2.54E-09 | 6.70E-08 | 6.95E-08 | 2.45E-09 |
| 562422 | 1.04E-09 | 3.09E-08 | 3.19E-08 | 1.12E-09 |
| 750001 | 4.27E-10 | 1.42E-08 | 1.47E-08 | 5.05E-10 |
| 1000143 | 1.75E-10 | 6.54E-09 | 6.71E-09 | 2.27E-10 |

**FIGURE CAPTIONS**

**Figure 1:** The He and C nuclei intensities below 1.5 GeV/nuc measured at V1 beyond the heliopause. C intensities are times 35. Calculations in this paper using a truncated exponential LBM with Tr=0.04 for He and 0.12 for C are shown as solid black lines. Large open circle shows normalization between He and Cx35 spectra. Dashed blue lines are GALPROP-DR calculations for He and C nuclei from Cummings, et al., 2016.

**Figure 2:** Observations and measurements of the He/C ratio between 3 MeV/nuc and $10^3$ GeV/nuc. Errors on individual GeV/nuc Voyager measurements are $\pm 5\%$. Errors on AMS-2 measurements are less than $\pm 2\%$. Black curve, labelled $P_0$=0.562*, is for a truncated exponential PLD with mean path length = $\lambda$=20.6 $\beta$ $P^{-0.45}$ above $P_0$=0.562 and with truncation parameters=0.04 for He and 0.12 for C. The curve labelled $P_0$=0.562 is for a simple LBM with a PLD=exponential at all path lengths for $P_0$=0.562 GV. Dashed blue line is GALPROP calculation of the He/C ratio from Cummings, et al., 2016.

**Figure 3:** A comparison of the PLD at small mean path lengths for an exponential PLD=8 g/cm$^2$ and for a PLD truncated with $\lambda_2$, = 0.04 and 0.12, showing a deficiency of short path lengths, $\lambda$, in the truncated models.

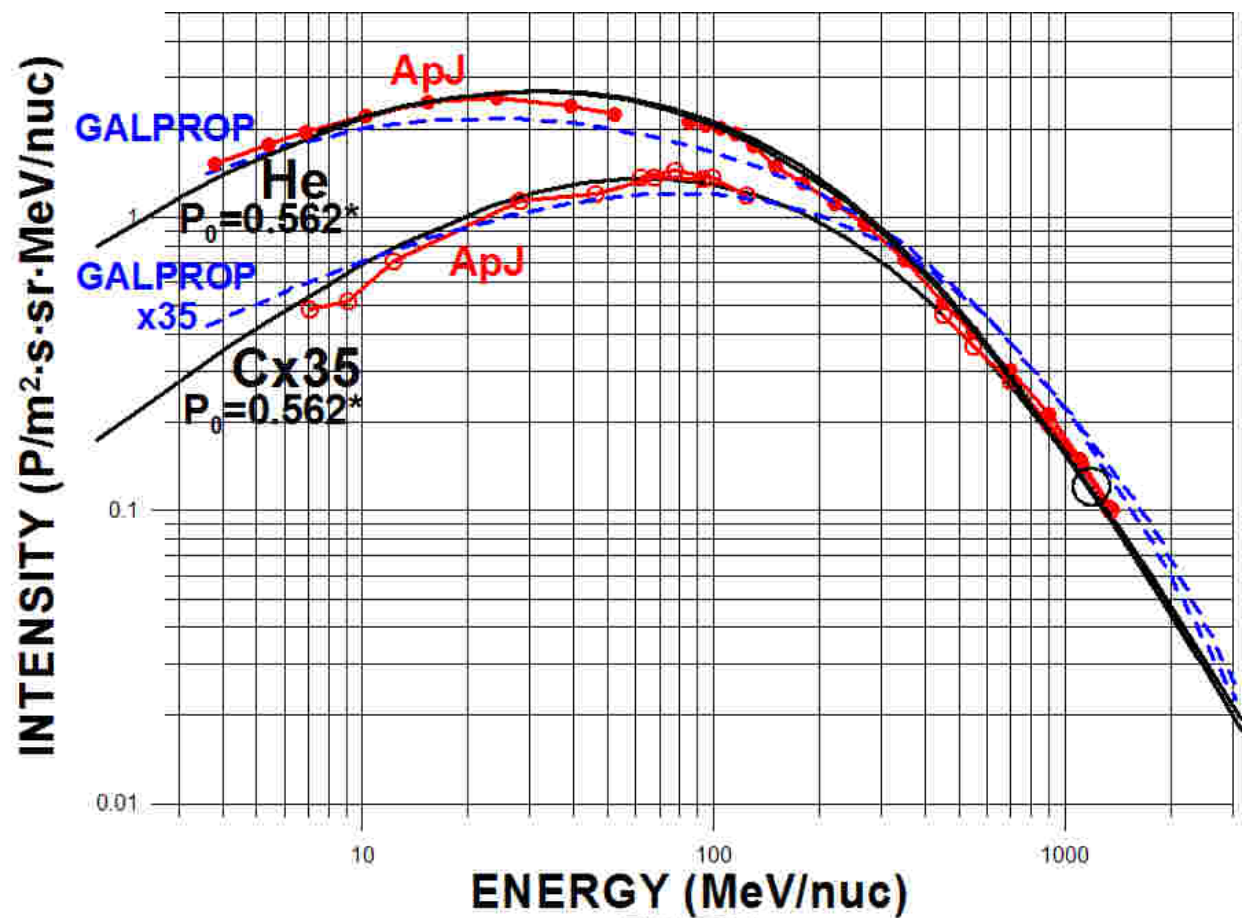

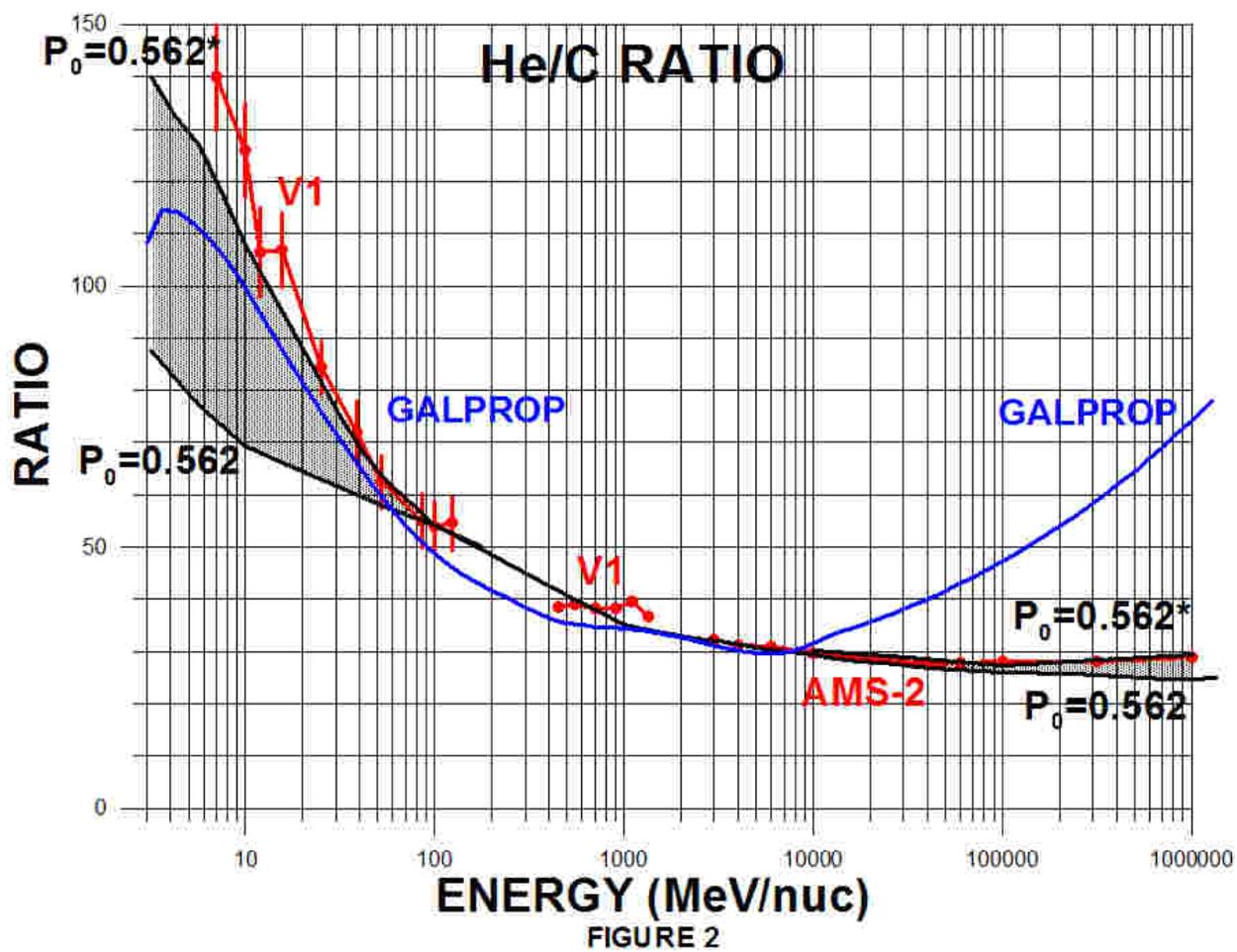

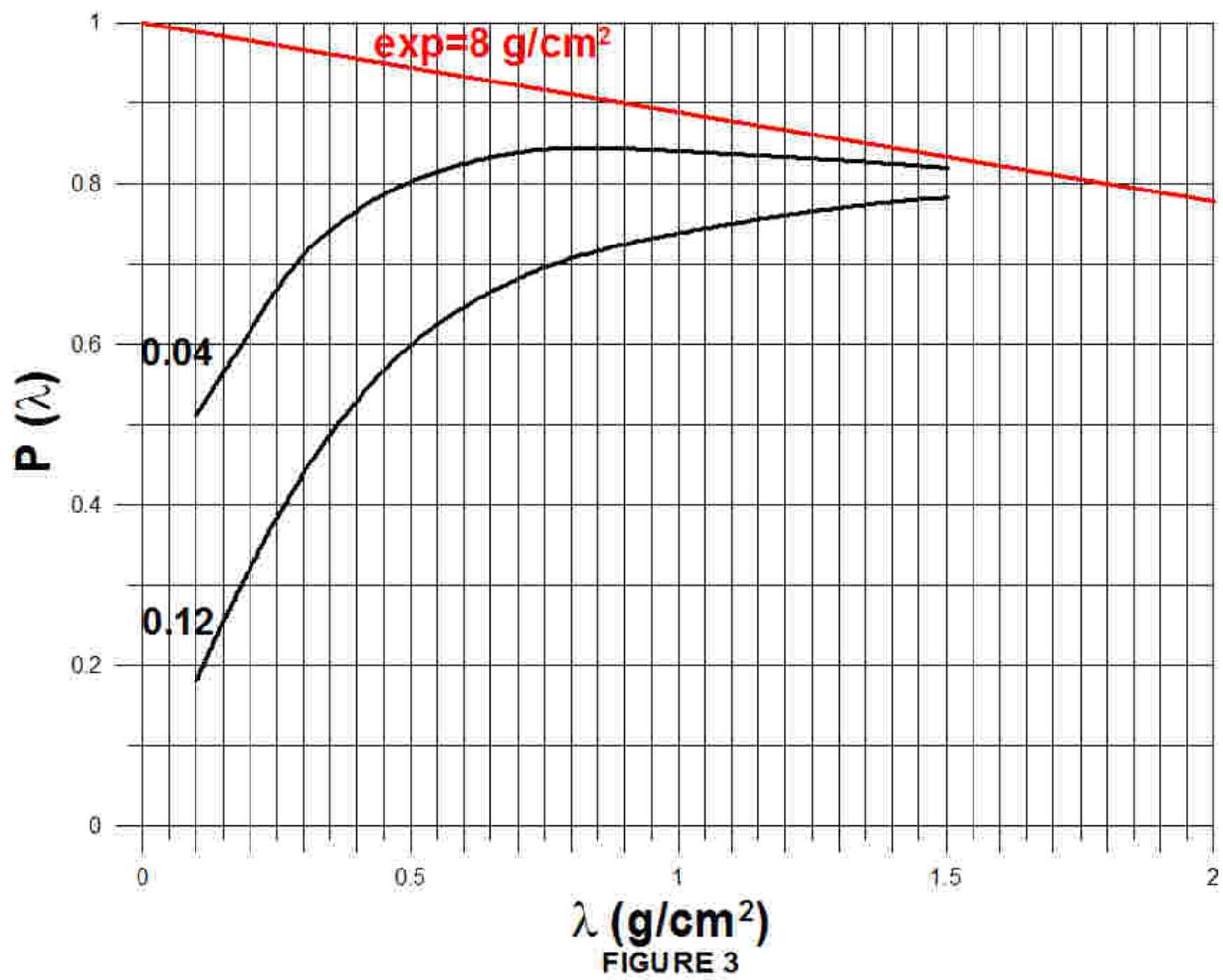

FIGURE 3